\documentclass[aps,prl,reprint,groupedaddress]{revtex4-2}

\usepackage{amsmath,amssymb}
\usepackage{bm,color}

\usepackage{graphicx,graphics}

\newcommand{\Fref}[1]{Fig.\ref{#1}}
\newcommand{\eref}[1]{eq.(\ref{#1})}

\begin{document}

\title{Dimensional reduction in evolving spin-glass model:\\
correlation of phenotypic responses to environmental and mutational changes}

\author{Ayaka Sakata}
\email[]{ayaka@ism.ac.jp}
\affiliation{Department of Statistical Inference \& Mathematics, Institute of Statistical Mathematics, 
10-3 Midori-cho, Tachikawa, Tokyo 190-8562,Japan}
\author{Kunihiko Kaneko}
\affiliation{
Center for Complex Systems Biology, Universal Biology Institute, University of Tokyo, 3-8-1 Komaba, Meguro-ku, Tokyo 153-8902, Japan}

\date{\today}

\begin{abstract}
The evolution of high-dimensional phenotypes is investigated using a statistical physics model consists of
interacting spins, in which 
phenotypes, genotypes, and environments are represented by spin configurations, interaction matrices, and external fields, respectively. We found that phenotypic changes upon diverse environmental change and genetic variation are highly correlated across all spins, consistent with recent experimental observations of biological systems. The dimension reduction in phenotypic changes is shown to be a result of the evolution of the robustness to thermal noise, achieved at the replica symmetric phase.
\end{abstract}


\maketitle

Biological systems generally consist of a huge number of components. Biomolecules (proteins) consist of a large number of monomers (amino acids), whereas cells consist of a variety of proteins, mRNAs, and other chemicals. Despite such high-dimensionality, however, there is growing evidence that the responses of phenotypes 
to external changes are often restricted to a low-dimensional subspace. 
 
For instance, the concentrations of a huge variety of components such as mRNAs and proteins have been recently measured against a variety of environmental stresses. The changes in  the (logarithmic) concentrations of mRNAs or proteins are found to be correlated \cite{Barkai,MA1,Bahler} or proportional \cite{Matsumoto,KK-PRX,Heinemann} across all components, 
against a variety of environmental stresses. This global proportionality suggests that phenotypic changes against environmental perturbations are constrained along a one- or low-dimensional manifold, a manifestation of a drastic dimension reduction from the high-dimensional composition space\cite{CFKK-PRE,KKCF-AnnrevBP}. 
Indeed, such dimension reduction would be rather universal in biological systems, as reported in studies of protein dynamics \cite{Tlusty}, ecological systems \cite{Leibler}, and neural learning dynamics \cite{learning}.
This global proportional change is also extended to the evolutionary dimension.
Changes in each concentration upon genetic mutation and those upon environmental perturbations are also highly correlated \cite{Geno-Pheno,Pancaldi,CFKK-Interface,Horinouchi1,Horinouchi2}.
It has been recently conjectured that such dimension reduction is a consequence of the evolution to achieve functional phenotypes that are robust to perturbations. Although some evolution simulations of catalytic-reaction networks support this conjecture\cite{CFKK-PRE,Sato-KK}, thus far, the concept remains an intuitive sketch, and an underlying mathematical structure remains elusive.
 
At this moment, a statistical physics approach would be useful to address the question of if and how the dimension reduction evolves.
Previously, we  studied a statistical physics model of spins, whose stochastic change is governed by a Hamiltonian that includes the two-body spin-spin interaction $J_{i,j}$ under thermal noise, specified by the temperature \cite{Sakata,Sakata2}.
In the model, the following correspondences are taken:  
phenotypes $\rightarrow$ spin configurations \{$S_i$\}; rule to shape the phenotype $\rightarrow$ Hamiltonian for spin-spin interaction $H=-\sum_{i,j}J_{ij}S_iS_j $;
environmental condition $\rightarrow$ external field $h_i$
to each spin in the Hamiltonian.
The evolution process is introduced by the ``mutation'' in $J_{i,j}$ and a selection according to the fitness defined from the spin configuration. 
By evolving the Hamiltonian under a certain temperature, we
have previously demonstrated the evolution of Hamiltonians to shape phenotypes to be robust to perturbations at an intermediate temperature 
corresponding to replica-symmetric (RS) phase,
whereas replica symmetry breaking (RSB) at lower temperature leads to rugged energy landscape and a non-robust phenotype. Still, the dimension reduction and its relationship with these phases was not investigated, which is one of the main focuses of the present Letter.

By taking advantage of this spin model and evolving it under a certain temperature, one can investigate if the dimension reduction
in phenotypic changes, as 
observed in biological systems,
is formulated and understood in terms of statistical physics. Specifically, we focus on the following questions: 
(i) Are high-dimensional phenotypic changes against various
environmental changes correlated? (ii) Are the changes induced by environmental and genetic changes correlated? 
(iii) If the above two correlations are observed, are they a result of dimension reduction from a high-dimensional phenotypic space, shaped by evolution?
(iv) Finally, within what range of temperature are the above questions answered affirmatively? In other words, 
is the appropriate 
noise relevant to the evolution of dimension reduction? 
By answering these questions, we will elucidate the origin of dimension reduction in terms of statistical physics, in possible relationship with RS/RSB.

Now, we define a spin-statistical physics model for phenotypic evolution, in which the phenotype is denoted by spins $\bm{S}=[S_1,\cdots,S_N]\in\{-1,+1\}^N$.
The dynamics of the spins are given by the stochastic dynamics, prescribed by the Hamiltonian $H$ as
\begin{align}
H(\bm{S}|\bm{J})=-\frac{1}{2}\bm{S}^{\mathrm{T}}\bm{J}\bm{S},
\label{eq:H}
\end{align}
where superscript $\mathrm{T}$ denotes the transpose,
and $\bm{J}\in\mathbb{R}^{N\times N}$ is a symmetric matrix
whose diagonal components are zero. 
With this Hamiltonian, the spin dynamics with discrete time $t$
is given by the transition probability
\begin{align}
\mathrm{Pr}[\bm{S}^{(t)}\to\bm{S}^{(t+1)}|\bm{J}]=\min\{e^{-\beta\Delta H(\bm{S}^{(t)},\bm{S}^{(t+1)}|\bm{J})},1\},
\label{eq:S_transition}
\end{align}
where $\bm{S}^{(t)}$ is the phenotype at step $t$, and 
$\Delta H(\bm{S},\bm{S}^\prime|\bm{J})\equiv H(\bm{S}^\prime|\bm{J})-H(\bm{S}|\bm{J})$.
Here, $\bm{S}^{(t+1)}$ differs from $\bm{S}^{(t)}$ only by a single site, hence spin configuration is asynchronously updated.
The inverse temperature $\beta=T^{-1}$ describes the stochasticity of the phenotype expression process.
The elements of the interaction matrix are chosen as $J_{ij}\in\Omega_J~(i\neq j)$
with $\Omega_J=\{-1\slash\sqrt{N},0,1\slash\sqrt{N}\}$,
and $J_{ii}=0$ $(i=1,\cdots,N)$.
This matrix represents the genotype, which evolves over generations, as will be described later.

The fitness is generally given as a function of phenotypes, i.e., the spin configuration.
Here, we assume that 
a part of the spins, named targets $i\in {\cal T}$, 
contributes to the fitness, 
such as the active site residues of protein. 
As more of the target spins have the same value $+1$
or $-1$, the fitness $\psi(\bm{J})$ is higher,
as defined as 
\begin{align}
\psi(\bm{J})=\overline{|{m_{\cal T}}|},~~~
m_{\cal T}=\frac{1}{N_T}\sum_{i\in {\cal T}}S_i
\end{align}
where $N_T$ is the size of ${\cal T}$, and
$\overline{\cdots}$ denotes the average over the trajectories of the 
phenotype expression dynamics, which depend on genotype $\bm{J}$.

The evolution to select genotypes with higher fitness is represented by
the following stochastic update rule with discrete time,
\begin{align}
\mathrm{Pr}[\bm{J}^{(g)}\to\bm{J}^{(g+1)}]=\min\{e^{\beta_J\Delta\psi(\bm{J}^{(g)},\bm{J}^{(g+1)})},1\},
\end{align}
where $\Delta\psi(\bm{J}^\prime,\bm{J})=\psi(\bm{J}^\prime)-\psi(\bm{J})$.
The parameter $\beta_J=T_J^{-1}$ represents the selection pressure;
as $T_J$ decreases, 
the genotype with higher
fitness survives to the next generation with 
high probability.

We mainly describe the results for
$N=100$ and $\rho\equiv N_T\slash N=0.05$, unless otherwise mentioned.
For the phenotype dynamics \eref{eq:S_transition}, we adopt the 
Markov Chain Monte Carlo (MCMC) method with detailed balance condition.
After a sufficient number of updates, the distribution of $\bm{S}$ is expected to converge to 
the equilibrium distribution,
$P(\bm{S})\propto\exp(-\beta H(\bm{S}|\bm{J}))$, for a given genotype.
We numerically calculated the thermal average
over $t_{\mathrm{f}}=2\times10^4$ MC steps,
after discarding the 
initial $t_{\mathrm{i}}=10^4$ steps.

At each generation $g$,
The candidates of genotype $\bm{J}^{(g+1)}$ are generated by 
introducing the mutations with probability $p_\mu=0.05$,
hence $\bm{J}^{(g+1)}$
differs from $\bm{J}^{(g+1)}$ by
$0.05\times N(N-1)\slash 2$ components.
\footnote{We have confirmed that $p_\mu=0.05$ does not change the ensemble ${\cal J}(T)$ that obtained by the asynchronous update of $\bm{J}$.}
The values of $J_{ij}$ ($i\neq j$) change into one of the components in $\Omega_J\backslash J_{ij}$
with equal probability,
where $A\backslash a$ denotes the members of $A$, excluding $a$.
We numerically update genotypes over generation
$g_\mathrm{max}=10^5$ at $T_J=0.05$. 
\footnote{This choice of $T_J$ is appropriate to
investigate the $T$-dependence of evolved genotypes.
For the $T_J$-dependence, see also \cite{Sakata2}}.
Without a loss of generality,
hereafter, we set the target sites as ${\cal T}=\{1,\cdots,N_T\}$.
We numerically obtain 100 genotypes 
evolved at $\rho$ and $T$ with different initial conditions, 
and the set is denoted as ${\cal J}(T)$.

\begin{figure}
\begin{minipage}{0.49\hsize}
\centering
\includegraphics[width=1.7in]{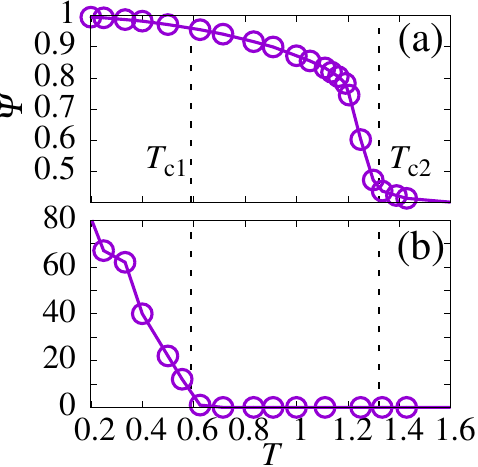}
\end{minipage}
\begin{minipage}{0.49\hsize}
\centering
\includegraphics[width=1.7in]{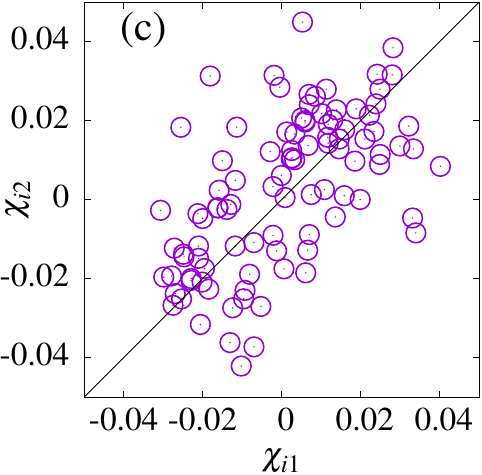}
\end{minipage}
\begin{minipage}{0.49\hsize}
\centering
\includegraphics[width=1.7in]{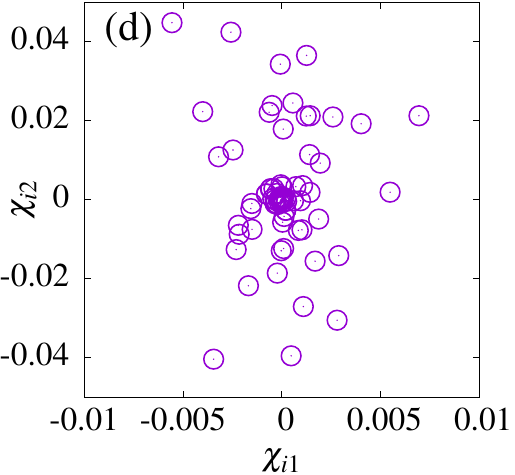}
\end{minipage}
\begin{minipage}{0.49\hsize}
\centering
\includegraphics[width=1.6in]{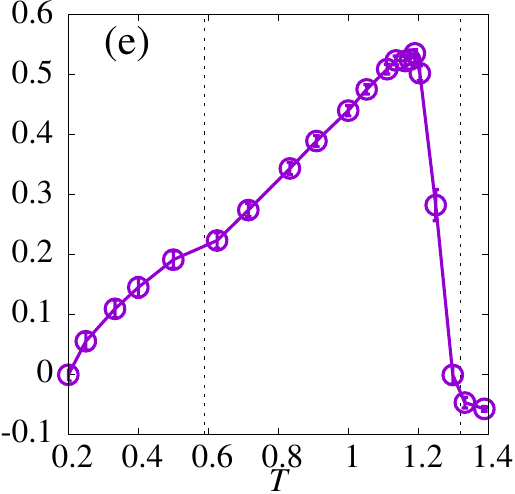}
\end{minipage}

\caption{(a) $T$-dependence of the averaged 
fitness. (b) Fraction of matrices ${\cal J}(T)$ in which the BP algorithm does not converge within $10^5$ steps. 
For (a) and (b), the vertical dashed lines indicate the phase transition temperatures.
(c)-(d) Scatter plots of $\chi_{i1}$ and $\chi_{i2}$ at (c) $T=1$  
and (d) $T=0.4$. 
The slope of the diagonal line in (c) is 1. 
(e) $T$-dependence of the 
averaged correlation coefficients between 
$\{\chi_{i1}\}\cdots\{\chi_{i,N_T}\}$.
The statistical errors over ${\cal J}(T)$ is smaller than the point size in the figure for
all $T$ region, and errorbars are not discernible.
}
\label{fig:fitness}
\end{figure}

First, we present the existence of three phases that depend on $T$ \cite{Sakata,Sakata2}.
\Fref{fig:fitness}(a) shows
the temperature dependence of the averaged 
fitness over ${\cal J}(T)$.
At $T\geq T_{c2}$,
the fitness value approaches
0.375 as $T$ increases, which is the level expected by the
random spin configuration
\footnote{The target-spin configurations
here 3 possibilities; 
(i) 5 spins aligned, $|m_{\cal T}|=1$, with probability $2/32$,
(ii) 4 spins aligned, $|m_{\cal T}|=0.6$, with probability $10/32$,
(iii) 3 spins aligned, $|m_{\cal T}|=0.2$, with probability $20/32$.
The summation of these leads $\psi=3/8$.}. Hence,
the phase $T\geq T_{c2}$ is identified as paramagnetic phase.
The high-fitness phase is separated into two phases at $T=T_{c1}$,
The region at $T_{c1}\leq T< T_{c2}$ is RS phase,
as is characterized by the convergence of the
belief propagation (BP) algorithm \cite{Mezard-Montanari}
\footnote{
In the fully connected system, the stability condition of BP algorithm agrees with the validity of RS assumption,
which is known as de Almeida-Thouless (AT) instability \cite{Kabashima,AT}.
BP algorithm is generally adopted as a numerical method to judge the
RSB transition for not-fully connected system, where analytical
derivation of AT instability is not available}.
The fitted state is reached fast enough and is robust to noise and mutation.
As shown in \Fref{fig:fitness}(b),
the fraction of $\bm{J}\in{\cal J}(T)$,
in which the BP algorithm
does not converge within $10^5$ steps,
increases from zero at $T_{c1}$.
Hence, the phase at $T<T_{c1}$ correspond to the RSB phases, as characterized by the rugged energy landscape (see also \cite{Sakata}).

Now, we discuss if the response to different environmental conditions is correlated or not, depending on the phase.
Hereafter, we study the symmetry breaking local magnetization 
$\mu_i=\overline{\mathrm{sign}(m_{\cal T})S_i}$,
considering the $\mathrm{Z}_2$ symmetry
\footnote{The definition of $\mu_i$
is because of the numerical convenience.
Another definition such as $\mu_i=\overline{\mathrm{sign}(\sum_{j=1}^NS_j\slash N)S_i}$ does not change the results.}.
Under the infinitesimal external fields,
the difference between expression patterns
$\delta\mu_i^{(h)}(\bm{h},\bm{J};\delta\bm{h})
\equiv\mu_i(\bm{h}+\delta\bm{h},\bm{J})-\mu_i(\bm{h},\bm{J})$
is expanded as
\begin{align}
\delta\mu_i^{(h)}(\bm{h},\bm{J};\delta\bm{h})\sim\sum_j\chi_{ij}(\bm{h},\bm{J})\delta h_j,\label{eq:dmu_h}
\end{align}
where $\chi_{ij}(\bm{h},\bm{J})=\partial \mu_i(\bm{h},\bm{J})\slash \partial h_j$ is the susceptibility.
We regard \eref{eq:dmu_h} as the response of the $i$-th component
to the additional external field,
for a system with genotype $\bm{J}$ subject to 
external field $\bm{h}$.
For simplicity, we consider the case that an external field 
$\delta\bm{h}_i$, whose $i$-th component is $\delta h (\neq 0)$,
otherwise 0, is applied to the system at $\bm{h}=\bm{0}$.
The first-order response of the $j$-th component to $\delta\bm{h}_i$ is $\chi_{ji}(\bm{0},\bm{J})$.
At the equilibrium, $\chi_{ij}(\bm{h},\bm{J})=\beta(\langle S_iS_j\rangle_{\bm{h}}-\langle S_i\rangle_{\bm{h}} \langle S_j\rangle_{\bm{h}})$ holds,
where $\langle\cdot\rangle_{\bm{h}}$ means the average
according to the
equilibrium distribution under the external field $\bm{h}$; $P(\bm{S})\propto\exp(-\beta H(\bm{S}|\bm{J})+\beta\bm{h}^{\mathrm{T}}\bm{S})$.
We numerically compute $\chi_{ij}$ by MCMC simulation
as $\chi_{ij}=\beta(\overline{S_iS_j}-\mu_i\mu_j)$.
\Fref{fig:fitness} shows the scatter plots of $\chi_{i1}$
and $\chi_{i2}$ under one realized genotype for $i\geq 3$
at (c) $T=1$ (RS) and (d) $T=0.4$ (RSB).
Their correlation coefficients are (c) 0.59,
and (d) -0.035, respectively.
Here, we ignore the responses of $\mu_1$ and $\mu_2$
to remove the trivial strong response directly to
$\delta\bm{h}_1$ and $\delta\bm{h}_2$ itself.
In \Fref{fig:fitness}(e), $T$-dependence of 
the correlation coefficient between
$\{\chi_{i1}\}\cdots\{\chi_{i,N_T}\}$ 
is shown, which is averaged over ${\cal J}(T)$.
The correlation between the responses to external fields $\delta\bm{h}_i$
($i\in{\cal T}$) is discernible
in the RS phase
\footnote{The correlation between responses to 
$\delta\bm{h}_i$ for $i> N_T$
are small compared with those of the target spins.
As will be discussed later, this is a consequence of the evolution
under the fitness defined on the target spins.}.

\begin{figure}
\begin{minipage}{0.49\hsize}
\centering
\includegraphics[width=1.78in]{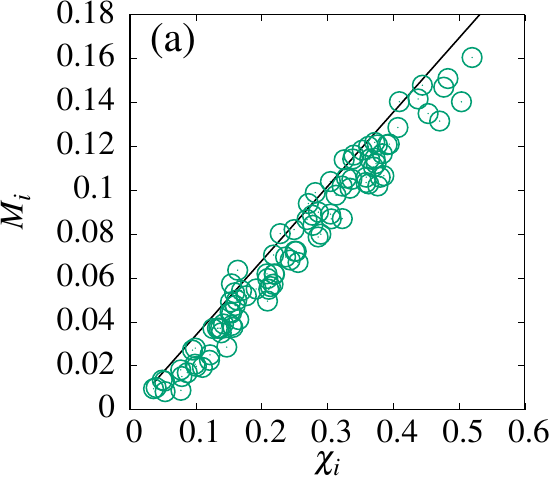}
\end{minipage}
\begin{minipage}{0.49\hsize}
\centering
\includegraphics[width=1.69in]{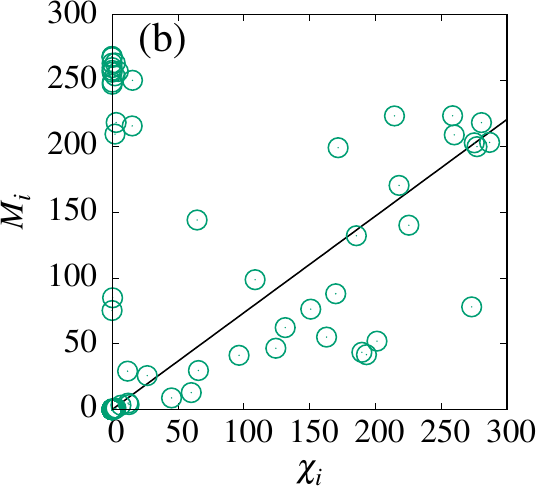}
\end{minipage}
\caption{Relationship between $\{\chi_i\}$
and $\{{\cal M}_i\}$ for one evolved genotype $\bm{J}$ at 
(a) $T=1$ (RS) and (b) $T=0.2$ (RSB). These behaviors are commonly observed for any evolved $\bm{J}(T)$.}
\label{fig:chi_vs_mu}
\end{figure}

Next, we study the correlation between responses to the environment, 
$\delta\mu_i^{(\bm{h})}$, and those to genetic changes,
$\delta\mu_i^{(J)}(\bm{J};\delta\bm{J})\equiv
\mu_i(\bm{0},\bm{J}+\delta\bm{J})-\mu_i(\bm{0},\bm{J})$,
expanded as
\begin{align}
\delta\mu_i^{(J)}(\bm{J};\delta\bm{J})
&\sim\sum_{jk}{\cal M}_{i,j<k}(\bm{J})\delta J_{jk},\label{eq:dmu_J}
\end{align}
where ${\cal M}_{i,jk}=\partial\mu_i(\bm{J})\slash J_{jk}$,
which corresponds to 
$\beta(\langle S_iS_jS_k\rangle-\langle S_i\rangle\langle S_jS_k\rangle)$
at the equilibrium.
For the comparison between $\delta\mu_i^{(h)}$ and $\delta\mu_i^{(J)}(\bm{J})$,
we assume that the components of 
$\delta\bm{h}$ and $\delta\bm{J}$
independently follow a Gaussian distribution
with mean 0 and variance $\epsilon$ for $\delta\bm{h}$,
and variance $\epsilon\slash\sqrt{N}$ for $\delta\bm{J}$,
respectively.
The expected squared responses are given by
\begin{align}
E_{\delta\bm{h}}\left[{\delta\mu_i^{(h)}}^2(\bm{h},\bm{J};\delta\bm{h})\right]&\simeq \epsilon^2\chi_i(\bm{h},\bm{J})\\
E_{\delta \bm{J}}\left[{\delta\mu_i^{(J)}}^2(\bm{J};\delta\bm{J})\right]&\simeq \epsilon^2{\cal M}_i(\bm{J}),
\end{align}
where $E_{\delta\bm{h}}[\cdot]$ and $E_{\delta \bm{J}}[\cdot]$ denote the average over $\delta\bm{h}$ and $\delta\bm{J}$,
respectively,
and $\chi_{i}=\sum_{j\neq i}\chi_{ij}^2$,
and ${\cal M}_i=N^{-1}\sum_{j<k,j,k\neq i}{\cal M}_{i,jk}^2$.
The quantities $\chi_{i}(\bm{h},\bm{J})$ and 
${\cal M}_i(\bm{J})$ correspond to the spin-glass susceptibility and ``susceptibility to interaction matrix,''
and indicate
the sensitivity of the $i$-th component
to the external field and mutation,
respectively.
\Fref{fig:chi_vs_mu} shows the scatter plot between ${\cal M}_i(\bm{J})$ and 
$\chi_i(\bm{0},\bm{J})$
for genotype $\bm{J}\in{\cal J}(T)$
at (a) $T=1$ (RS) and (b) $T=0.2$ (RSB).
A linear relationship between $\chi_i$ and ${\cal M}_i$ arises 
in the RS phase.

\begin{figure}
\begin{minipage}{0.49\hsize}
\centering
\includegraphics[width=1.5in]{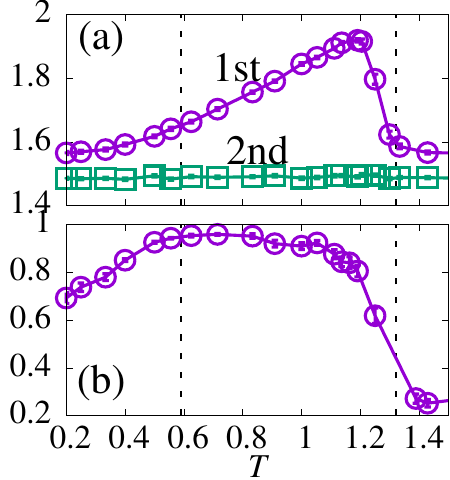}
\end{minipage}
\begin{minipage}{0.49\hsize}
\includegraphics[width=1.8in]{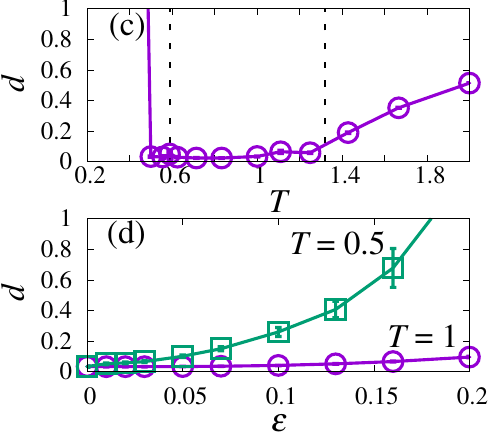}
\end{minipage}
\caption{$T$-dependence of (a) averaged the first and second eigenvalues of $\bm{J}$, 
(b) correlation coefficient between $\mathrm{arc}\tanh(\mu_i)$
and $\xi_i^1$,
and (c) the averaged $d$.
Vertical dashed line denotes transition temperatures.
(d) $\epsilon$-dependence of $d$ for 
$T=1$ (RS) and $T=0.5$ (RSB).}
\label{fig:eigenvalues}
\end{figure}

These numerical simulations indicate that
the evolution under thermal fluctuation 
that leads to the RS phase induces 
the correlations between the responses.
To understand the emergence of the correlation,
we decompose the evolved genotypes into eigenvalues and eigenvectors
as $\bm{J}=\bm{\Xi}\bm{\mathrm{\Lambda}}\bm{\Xi}^{\mathrm T}$,
where $\bm{\mathrm{\Lambda}}\in\mathbb{R}^{N\times N}$
is a diagonal matrix consisting of eigenvalues 
$\mathrm{\Lambda}_{ii}=\lambda_i$ $(\lambda_1\geq\lambda_2\geq\cdots\geq\lambda_N)$,
and $\bm{\Xi}=[\bm{\xi}^1,\cdots,\bm{\xi}^N]\in\mathbb{R}^{N\times N}$
is the set of corresponding eigenvectors.
\Fref{fig:eigenvalues}(a) shows the averaged values 
of the first and second eigenvalues over $\bm{J}\in{\cal J}(T)$. 
The first eigenvalue is much larger in the RS phase than 
those in the other phases. The evolutionary change of the
second eigenvalue is vanishingly small for any $T$.
This tendency is common for any $\lambda_i$ ($i\geq 2$).
Hence, the
dominancy of the first eigenmode is enforced
as a result of the evolution at $T_{c1}\leq T\leq T_{c2}$.

On the basis of the large contribution of the first eigenvalue in the RS phase,
we apply a 1-rank approximation of genotype
$\bm{J}\sim\eta_1\bm{\xi}^1{\bm{\xi}^1}^{\mathrm{T}}$.
By a straightforward calculation, 
the local magnetization is expressed as
\begin{align}
\mu_i=\tanh\left(\beta\eta_1\xi_{i}^1\sum_{k\neq i}\xi_{k}^1\mu_k+h_i\right),
\label{eq:mu_i_1rank}
\end{align}
at sufficiently large $N$. 
Therefore, when the first eigenmode is dominant,
the relationship $\xi_{i}^1\propto\mathrm{atanh}(\mu_i)$
should hold at $\bm{h}=\bm{0}$.
\Fref{fig:eigenvalues}(b) shows
the correlation coefficient between 
$\{\mathrm{atanh}(\mu_i)\}$ and $\{{\xi}_i^1\}$.
In the RS phase, the correlation coefficient approaches 1;
hence, $\xi_i^1\sim\mathrm{atanh}(\mu_i)$ is a reasonable approximation.
We note that the expression of $\bm{J}=\eta_1\bm{\xi}^1{\bm{\xi}^1}^{\mathrm{T}}$ 
is similar to those of the Mattis model \cite{Mattis,Amit},
which is the Hopfield model with a single embedded pattern \cite{Hopfield}.
The present embedded pattern, however, is
$\sqrt{\eta_1}\bm{\xi}^1$,
in contrast to a discrete vector with $\pm 1$
in the Mattis model.
For sufficiently small $\rho$,
the distribution of $\mu_i$ is almost random, and 
the embedded pattern after the evolution is 
a random pattern, except the target spins
\footnote{See Fig.1 of supplements for
$\rho$-dependence of the embedded pattern}.
Even though the approximate estimate by Mattis-type model is used here, the evolved genotypes in RS phase do not perfectly agree with it: Indeed, eigenmodes other than the first mode remain, which induces frustration between non-target spins \cite{Sakata}. This hampers the correlation between responses
of non-target spins.

Last, we show that the dominancy of the first eigenmode of genotype induces a correlation between the responses to environmental and genetic changes,
as observed in the RS phase.
From \eref{eq:mu_i_1rank},
we obtain the expression for susceptibility under the 1-rank approximation
\begin{align}
    \chi_{ij}=v_i\left(\delta_{ij}+\eta_1\xi_{i}^1\sum_{k\neq i}\xi_{k}^1\chi_{kj}\right),
    \label{eq:chi_1rank}
\end{align}
where $v_i=\beta(1-\mu_i^2)$ and $\delta_{ij}$ is Kronecker's
delta.
Because of the randomness of the embedded pattern,
it is reasonable to assume that $\chi_{ij}~(i\neq j)$
is 
sufficiently small; hence,
$\langle S_iS_j\rangle\sim\mu_i\mu_j$ holds.
Applying the equilibrium relationship 
${\cal M}_{i,jk}=\partial\langle S_jS_k\rangle\slash\partial h_i$, 
we obtain 
${\cal M}_{i,jk}\sim\chi_{ij}\mu_k+\mu_j\chi_{ik}$.
Because $\{\mu_i\}$ is expected to be randomly distributed,
${\cal M}_i=\sum_{jk}\chi_{ij}^2\mu_k^2$ holds,
neglecting the cross-term.
Setting $Q\equiv N^{-1}\sum_i\mu_i^2$, we obtain
\begin{align}
{\cal M}_i(\bm{J})=\chi_i(\bm{0},\bm{J})Q.
\label{eq:prop}
\end{align}
Hence, the proportionality between $\{\chi_i\}$
and $\{{\cal M}_i\}$ is a 
consequence of the dominance of the first eigenmode evolved in the RS phase,
i.e., the evolutionary dimensional reduction.
Here, notice that for the Mattis system,
\eref{eq:prop} itself holds but ${\cal M}_i$
and $\chi_i$ are not distributed and take unique values 
over all $i$,
hence the
proportionality 
between distributed ${\cal M}_i$
and $\chi_i$
as in \Fref{fig:chi_vs_mu} is not observed. The distribution comes from the non-target spins in our model.

The relationship \eref{eq:prop}
is indicated by the solid line in \Fref{fig:chi_vs_mu}.
We quantify the deviation of the observed
$\chi$-${\cal M}$ relationship 
from the theoretical line \eref{eq:prop},
by the normalized mean squared error
$d=\sum_i({\cal M}_i-Q\chi_i)^2\slash\sum_i{\cal M}_i^2$.
\Fref{fig:eigenvalues}(c) shows the $T$-dependence of $d$ 
averaged over ${\cal J}(T)$.
In the RS phase, $d$ is close to 0; hence,
\eref{eq:prop} holds with high accuracy,
which is a result of the emergence of the dominant first eigenmodes,
accompanied by randomness in the non-target spins.

When $T(<T_{c2})$ is close to the RS-RSB boundary,
$d$ is close to 0, as with the RS phase.
The difference between the RS and RSB phase 
is clear for finite $\bm{h}$ and $\Delta\bm{J}$,
which is a deviation of $\bm{J}$ from ${\cal J}(T)$.
We randomly generate $\bm{h}\sim{\cal N}(\bm{0},\epsilon^2\bm{I})$
and symmetric $\Delta\bm{J}$, where 
$\Delta J_{ij}\sim{\cal N}(0,\epsilon^2\slash N)$,
and $\Delta J_{ii}=0~\forall i$.
We quantify the relationship between 
${\chi}_i(\bm{h},\bm{J})$ and 
${\cal M}_{i}(\bm{J}+\Delta\bm{J})$ using $d$.
\Fref{fig:eigenvalues}(d) shows $\epsilon$-dependence of
the averaged $d$ over ${\cal J}(T)$ and 100 samples of 
$\bm{h}$ and $\Delta\bm{J}$
for $T=1$ (RS) and $T=0.5$ (RSB).
In the RSB phase,
$d$ increases faster than it does in RS phase,
even when $d$ at $\epsilon=0$ is close to zero.
This robustness of the proportionality is also
a consequence of the dominant first eigenmode
\footnote{
See Supplement Fig.2,
for $\epsilon$-dependence of $d$
over different values of target ratio $\rho$ at $T=1$.
Although a strong correlation at $\epsilon=0$
is observed for any $\rho$,
the relationship is not robust to noise $\epsilon$
as $\rho$ increases. The existence of redundant spins other than targets is relevant to robustness and dimension reduction.}.

The proportionality between $\{\chi_{ij}\}$ and $\{\chi_{ik}\}$~($j,k\leq N_T$, $j\neq k$),
shown in \Fref{fig:fitness}(b),
is also a consequence of the dominant first eigenmode.
From \eref{eq:chi_1rank}, the leading term of susceptibility is 
$\chi_{ij}=\eta_1\xi_{i}^1\xi_{j}^1v_iv_j$ ($i\neq j$);
hence, $\chi_{ij}\slash\chi_{ik}={\xi_j^1v_j}\slash(\xi_k^1v_k)$.
In the RS phase, both $v_j$ and 
$\xi_{j}^1$ are functions of $\mu_j$;
hence, $\chi_{ij}\slash\chi_{ik}\sim 1$ holds when $\mu_j\simeq\mu_k$.
This is the origin of the
linear relationship between $\{\chi_{ij}\}$ and $\{\chi_{ik}\}$
\footnote{Approximations \eref{eq:mu_i_1rank} and 
\eref{eq:chi_1rank} are relatively inaccurate for 
components with small local magnetization,
in the sense that they are sensitive to
the correction of first eigenmodes by taking higher modes into account.
Therefore, the correlations between $\{\chi_{ij}\}$ and $\{\chi_{ik}\}$ 
are observed for components whose local magnetizations are sufficiently large,
including target components.}.

In summary, we applied an evolving spin-statistical physics model, 
representing phenotypes, genotypes, and the environment by spin
configuration, interaction matrix, and the external field,
respectively, and have answered the questions addressed 
at the beginning of this paper. 
(i) Correlated responses across different environmental changes are demonstrated by the correlation in susceptibilities $\chi_{ij}$ and $\chi_{i\ell}$ in the evolved genotypes at the RS phase.
(ii) Proportional responses to mutation and environmental changes are demonstrated by the proportionality between the 
``susceptibility to interaction matrix'' ${\cal M}_i$ and spin-glass susceptibility $\chi_i$. 
(iii) These proportional responses originate in the reduction of rank in the interaction matrix. 
(iv) Such dimension reduction and proportional changes are 
observed for the evolved genotypes at the RS phase, 
i.e., at an intermediate level of thermal noise.
The RS phase was also evolved 
in a fully-connected system, 
where the frustration around target spins 
is diminished, as termed as local Mattis state \cite{Sakata}. 
The current study demonstrates that such RS phase (in a sparse connection) 
shows the correlated responses of 
phenotypes to environment and mutation, 
with dimension reduction, 
as supported by the redundant 
degrees of freedom by non-target spins.

Hence, robustness of phenotypes to noise \cite{Wagner,KK-PLoS} is essential to the evolutionary dimension reduction, leading to the correlated responses in the high-dimensional phenotypes to different types of perturbations. Although the present statistical physics model is highly simplified, it gives a theoretical basis for dimension reduction in biological systems, in which robustness to noise is also essential.
In fact, the present model can be interpreted as the evolution of protein to have a certain function. 
The RS phase here correspond to the funnel structure in contrast to the spin-glass phase \cite{Saito}. 
Note that recent reports on protein dynamics suggest the existence of large collective motion, which may be a manifestation of dimension reduction \cite{Tlusty,Togashi,Revoir,Murugan}. 
The correspondence between noise and mutation responses is also consistent with the simulation \cite{Geno-Pheno} and experiments \cite{Ichihashi} 
of the evolution of t-RNA. 
Last, although dynamics at the cellular level are not represented by a Hamiltonian, the similarity between spin-glass dynamics and gene expression dynamics with mutual activation and inhibition is now well recognized \cite{Kauffman,Derrida,Wagner,Mjolsness,KK-PLoS}.
In these examples, correlated phenotypic responses as a result of dimension reduction 
are evolutionarily acquired
as in the RS phase in our model at an intermediate temperature.

In terms of statistical physics, the evolution to the RS phase under appropriate levels of noise should be considered, in which both higher fitness and robustness to noise are achieved with the dimension reduction. If the temperature is reduced, robustness in the phenotype is lost by RSB, even though a higher fitness state is reached after sufficient time steps of expression.
Here, we have studied the simplest fitness condition. For higher biological functions, the response to diverse environmental conditions, say, different target spin configurations 
upon the application of 
different external fields, may be required. The extension to such problems would be straightforward, in which the need for both robustness and plasticity may lead to dimension reduction with higher ranks.

\begin{acknowledgments}
The authors thank to Koji Hukushima and Yoshiyuki Kabashima
for helpful comments and discussions.
This research was partially supported by a Grant-in-Aid for Scientific Research (S) (15H05746) and (wakate) (19K20363) from the Japanese Society for the Promotion of Science (JSPS) and Grant-in-Aid for Scientific Research on Innovative Areas (17H06386) from the Ministry of Education, Culture, Sports, Science and Technology (MEXT) of Japan.
\end{acknowledgments}

\end{document}